\documentclass[preprint,preprintnumbers,amsmath,amssymb,showkeys,prb]{revtex4-1}

\usepackage{filecontents}
\usepackage{graphicx}
\usepackage{dcolumn}
\usepackage{bm}

\begin{document}

\title[]{Intrinsic piezoelectricity of  PZT }

\author{Desheng Fu$^{1,2,3*}$, Seiji  Sogen$^2$ and Hisao Suzuki$^4$}

\affiliation{$^1$Department of Electronics and Materials Science,
Faculty of Engineering, Shizuoka University, 3-5-1 Johoku, Chuo-ku,
Hamamatsu 432-8561, Japan.\\
$^2$Department of Engineering, Graduate School of Integrated Science
and Technology, Shizuoka University, 3-5-1 Johoku, Chuo-ku,  Hamamatsu
432-8561, Japan.\\
$^3$Department of Optoelectronics and Nanostructure Science,
Graduate School of Science and Technology, Shizuoka University,
3-5-1 Johoku, Chuo-ku, Hamamatsu 432-8011, Japan.\\
$^4$ Research Institute of Electronics, Shizuoka University,3-5-1
 Johoku, Chuo-ku, Hamamatsu 432-8011, Japan.\\ 
 \rm{email:fu.tokusho@shizuoka.ac.jp} }




\begin{abstract}
An  unresolved  issue in the commonly used Pb(Zr$_{1-x}$Ti$_x$)O$_3$
(PZT) ceramics is understanding  the intrinsic  piezoelectric behaviors of its crystal
around the  morphotropic phase boundary (MPB). 
Here, we  demonstrate an  approach to  grow $c$-axis oriented tetragonal  PZT
 around MPB on  stainless steel SUS430,  allowing  us to
estimate the intrinsic piezoelectric and ferroelectric properties  of PZT
along its  polar axis. The piezoelectric coefficient $d_{33}$ and
spontaneous polarization $P_{\rm s}$ were found to  be 46.4 $\pm$ 4.4 pm/V, 88.7 $\pm$ 4.6
$\mu$C/cm$^2$, respectively, for $x=0.47$ close to MPB. These values align well  with 
 the predicted  values of $d_{33}=50\sim55$ pC/N and $P_{\rm
s}$=79 $\mu$C/cm$^2$  at room temperature from  the first-principles-derived approach. The
obtained $d_{33}$ is 4 times smaller than that of its ceramics,  indicating that the large piezoelectric response in the PZT ceramics around MPB  is primarily driven  by extrinsic effects rather than intrinsic ones.  In the technical application of PZT films, achieving a substantial piezoelectric response requires careful consideration of these extrinsic effects.

\end{abstract}

\keywords{PZT, morphotropic phase boundary (MPB), piezoelectric, ceramics, single crystal, oriented film, piezoelectric coefficient $d_{33}$ } 
\maketitle

\section{Introduction}
Pb(Zr$_{1-x}$Ti$_x$)O$_3$ (PZT) is an alloy of antiferroelectric
PbZrO$_3$  and ferroelectric PbTiO$_3$. It exhibits  a nearly vertical morphotropic
phase boundary (MPB) in the vicinity of  $x=0.47-0.50 $,  separating  the
 ferroelectric rhombohedral and tetragonal phases.  PZT ceramics display an exceptionally large piezoelectric response 
in the vicinity of the MPB.\cite{Berlincourt, Jaffe1, Jaffe2} This characteristic has established PZT ceramics as the primary workhorse in the realm of piezoelectric devices, with widespread applications in various fields, including ultrasonic imaging in medicine, sensors in automobiles, and  atomic positioning in science.\cite{Uchino}  

Beyond its evident technological significance, PZT  holds fundamental importance.
The MPB concept  derived from PZT has frequently guided the design of new 
materials with significant  physical property response. This has led to the successful discovery of high piezoelectric responses in ferroelectrics \cite{Park,Saito} and large magnetic responses in ferromagnets stemming from the MPB.\cite{Yang,Bergstrom} Clearly, comprehending the fundamental physics  involved in MPB remains  a critical concern in the  fields of  material sciences and condensed matter physics,\cite{George,Fu,Guo,Xu,Noheda1,Noheda2,Hall1,Hall2,Gharb,Hinterstein, Liu,Fan1,Fan2,Wang,Zhao,Gorfman,Sun,JLi,Asada,Otonicar} which may pave the way for designing  new functional materials with excellent physical properties.

Understanding the intrinsic piezoelectric behaviors of PZT single crystals near the MPB  have been a longstanding challenge.  Since  its discovery 
 seven decades ago,\cite{Shirane} substantial and continuous efforts have been dedicated to growing high-quality single crystals of PZT. \cite{Clarke,Phelan,Bokov,Roleder,Whatmore}   However, the unavailability of PZT single crystals with high quality has hindered the determination of their intrinsic piezoelectric properties near the MPB. Consequently,  theoretical approaches have been employed  to  address this issue. A semi-empirical simulation work on the basis of phenomenological Landau Devonshire theory  predicts that
a PZT single crystal would have a  $d_{33}$ value of 520 pm/V in the rhombohedral phase and  325
pm/V in the tetragonal phase along the polar axis near MPB.\cite{Du} The large $d_{33}$ observed in PZT
ceramics is attributed to be the response of a single crystal. In contrast, the  first-principles calculations  have found that the
$d_{33}$ values of a tetragonal PZT single crystal with $x=0.5$ is  3 times less than
the experimental value observed for ceramics at low
temperatures.\cite{Szabo, Bellaiche1} Furthermore,  
first-principles-derived approach has been  developed to study
the finite-temperature properties of PZT around  MPB, and the $d_{33}$
is predicted to be $50\sim55$ pC/N for
a tetragonal single crystal of PZT($x$=0.5) at room temperature,\cite{Bellaiche2} which  is approximately  6 times smaller than the one  derived from phenomenological Landau Devonshire theory.  The theoretical results derived from first-principles calculations and Landau theory calculations are fundamentally different. Therefore, even from a theoretical perspective, the intrinsic piezoelectric effects of PZT single crystal remain unclear. 

It is clear that the intrinsic piezoelectric response of PZT near MPB ultimately still requires experimental confirmation. The key to addressing this issue lies in the investigation of the linear piezoelectric effect in PZT along its polar axis, which is solely caused by  the lattice deformation induced by an electric field.
In this communication, we present a method to grow polar-axis oriented PZT to  identify the  intrinsic piezoelectric response near MPB.  The piezoelectric coefficient $d_{33}$ and  spontaneous polarization  $P_{\rm s }$ were experimentally found to be 46.4 $\pm$ 4.6 pm/V, 71.4 $\pm$ 11.3
$\mu$C/cm$^2$, respectively, in PZT with  $x=$0.47 close to MPB.   Our experimental results fully  support  the predictions based on  the first-principles  calculations by  L. Bellaiche et. al.\cite{Bellaiche1,Bellaiche2} and indicate that the high piezoelectric response in PZT ceramics is primarily driven by the  extrinsic effects rather than the intrinsic  effects of lattice response. 

\section{Experimental}

\begin{figure}
\includegraphics[width=8cm]{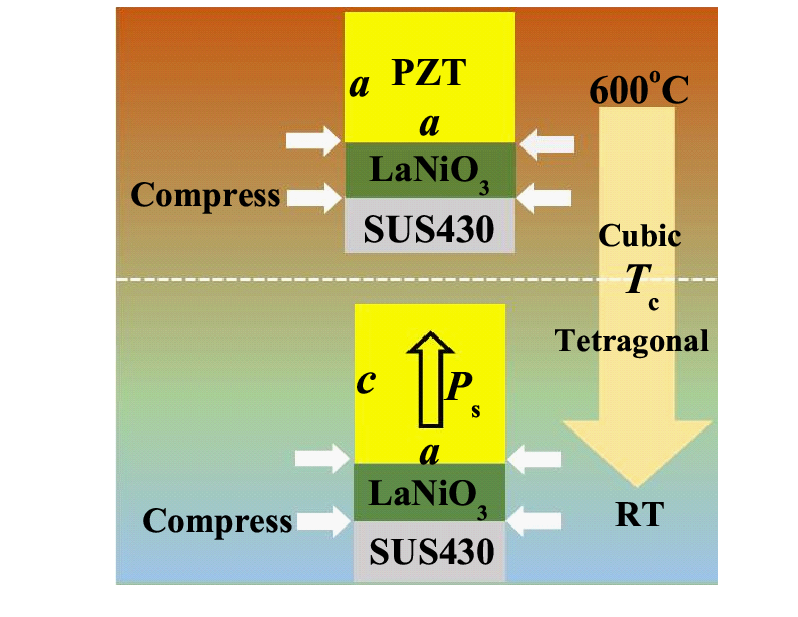}
\caption{\label{Fig1} Schematic diagram for  growing  $c$-axis
oriented Pb(Zr$_{0.53}$Ti$_{0.47})$O$_3$  using  compression from  thermal stress.}
\end{figure}

We utilized  chemical solution deposition\cite{Suzuki} to grow the tetragonal $c$-axis oriented  PZT with $x=$0.47,  close to MPB.  We have successfully developed an approach to grow  the $c$-axis oriented PZT through using a stainless steel SUS430 substrate and a  metallic LaNiO$_3$ (LNO) seed layer, as schematically shown in Fig.\ref{Fig1}.  LNO possesses  a pseudocubic perovskite structure with a lattice constant of $a= 3.838 {\AA}$,  comparable to that of  PZT with a perovskite structure. Moreover, LNO exhibits  the  least surface energy for the (100)-plane growth, enabling full (100)-plane  growth even on an amorphous  substrate.\cite{Miyake} Therefore,  the (100)-oriented LNO effectively serves  as  a seed layer to guide the  growth of  (100)- or (001)-oriented PZT with tetragonal structure on SUS430 substrate.  We further used  the significant  difference in  thermal expansion coefficient between SUS430 and PZT to achieve  (001)-plane  growth of tetragonal PZT, allowing  us to apply a compressive thermal stress  to PZT unit cell during cooling from the growth temperature.  This compressive thermal stress greatly promotes  the $c$-axis growth of tetragonal PZT as  the crystal transitions from the cubic to tetragonal phase during cooling.  A 0.24-$\mu$m-thick (100)-oriented  LNO was grown at 700$^\circ$C for 5 minutes in an oxygen atmosphere from  the LNO precursor solution prepared from La(NO$_3$)$_3$, Ni(CH$_3$COO)$_2$, and solvents of 2-methoxyethanol and 2-aminoethanol. The detailed process for preparing LNO is illustrated in Fig.S1 of the supporting information and can also be found in the previous report.\cite{Suzuki}  

To  prevent the diffusion of Cr from SUS430 to PZT, which  typically  reduces  the crystalline quality  and physical properties of PZT, a  0.3-$\mu$m-thick SiO$_2 $ buffer layer was placed between LNO and SUS430.  The detailed process for preparing  SiO$_2$ is shown in Fig.S2 of the supporting information. The SiO$_2$ precursor solution was prepared by dissolving Si(OC$_2$H$_5$)$_4$ in ethanol, followed by hydrolysis with water in the presence of hydrochloric acid as an acid catalyst. The purpose of the SiO$_2$ layer is to prevent chromium diffusion from the substrate, necessitating a dense structure. Using hydrochloric acid as a catalyst leads to the formation of precursor with linear or cross-linked molecular chain, enabling the creation of a dense film. The precursor solution was deposited on a SUS430 substrate by spin coating at 2500 rpm for 30 seconds. It was then dried at 150°C for 10 minutes,  pre-annealed at 350°C for 10 minutes to eliminate organic matter, and finally fired at 700°C for 10 minutes in an oxygen atmosphere to form the SiO$_2$ layer. This process was repeated three times to achieve the desired film thickness of the SiO$_2$ layer.

PZT was  subsequently  grown  on the LNO/SiO$_2$/SUS430 substrate at 650$^\circ$C in an oxygen atmosphere using a precursor solution prepared from Pb(OCOCH$_3$)$_{2}\cdot$3H$_2$O, Zr(OC$_3$H$_7$)$_4$, Ti(OCH(CH$_3$)$_2$)$_4$, and absolute ethanol as solvent. The detailed process for preparing  PZT films  are shown in Fig.S3 of the supporting information. To produce high-quality PZT thin films, we developed the precursor solution using metal alkoxides, employing a partial hydrolysis approach for Zr- and Ti-alkoxides with a chemical modification  using acetic acid and  controlling water addition for hydrolysis. This process led to the creation of Zr-O-Ti chemical bonds in the precursor solution, paving the way for exceptionally high-quality PZT films. PZT was grown  to a thickness of 0.3 $\sim$ 1.2  $\mu$m, which is sufficient to disregard  the impact  of two-dimensional  strain from the substrate  on the  overall physical properties of PZT. \cite{Schlom}

The crystal structure of the films was observed using a BRUKER AXS D8 ADVANCE X-ray diffractometer with Cu $k\alpha$ radiation. The surface of the films was examined using both field emission scanning electron microscopy (FE-SEM, JEOL JSM-7001F) and atomic force microscopy (AFM, SPI3800N, SII). Cross-sectional images of the film were obtained through transmission electron microscopy (TEM) using a JEM-2100F. The sample for TEM measurement was prepared by focused ion beam (FIB) with a JEOL JIB 4500.

Toyo ferroelectric tester system (FCE-3) in conjunction with  SII AFM  was used to  examined the polarization-switching and displacement behaviors of these $c$-axis oriented PZT films.  The samples, grown on 1 cm $\times$ 1 cm SUS430, were mounted on the AFM scanner using silver paste to firmly unite the scanner and SUS430 substrate, preventing the sample from bending. The measurements were conducted on capacitors with 60-$\mu$m-diameter top Pt electrodes  and conducting LNO bottom elecctrode at room temperature, at a frequency of 1 Hz. This low-frequency measurement allows for more complete switching of spontaneous polarization compared to the frequency of 1 kHz commonly reported in the literature.

\section{Results and discussion}
\begin{figure}
\includegraphics[width=8cm]{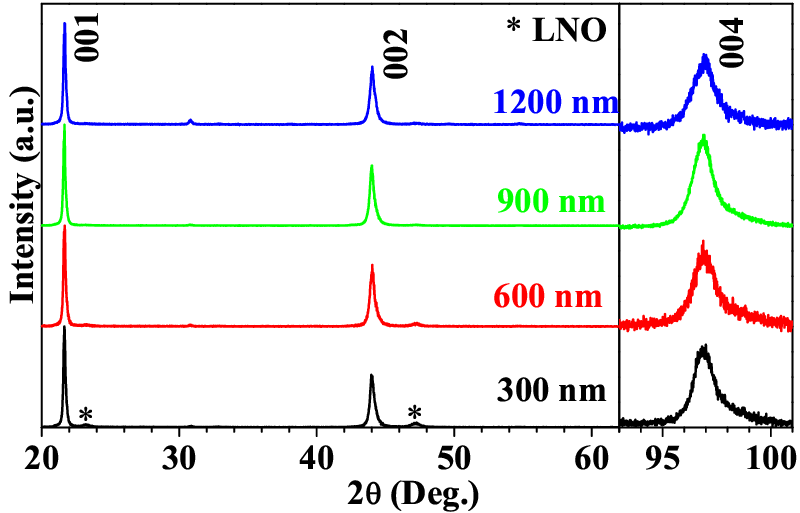}
\caption{\label{Fig2} X-ray diffraction patterns of $c$-axis
oriented Pb(Zr$_{0.53}$Ti$_{0.47})$O$_3$ grown on a SUS430
substrate. * marks the diffraction peaks positions of LNO.  }
\end{figure}

Figure \ref{Fig2} shows the X-ray diffraction patterns of the  PZT samples with thickness from 0.3 $\sim$ to 1.2  $\mu$m.  This result indicates that our approach successfully enabled the growth of $c$-axis (polar-axis) oriented PZT with a tetragonal structure on the SUS430 substrate. The lattice constant  was  determined to be 4.110 $\AA$ from the (004) diffraction peak observed in the XRD patterns, which is independent of the thickness of the PZT film in the range from 0.3 $\sim$ 1.2  $\mu$m. This value aligns excellently with the $c$-axis lattice constant of its bulk counterpart, which exhibits tetragonal symmetry with $c$=4.113$\AA$ and $a$=4.019$\AA$ at room temperature.\cite{Shirane} The independence of the lattice constant with film thickness and the agreement of the lattice constant between the sample and its bulk further verify that the grown PZT lattice has been fully relaxed from substrate strain. Consequently, the overall physical properties of the sample can be regarded as those of the bulk material. 

\begin{figure}
\includegraphics[width=8cm]{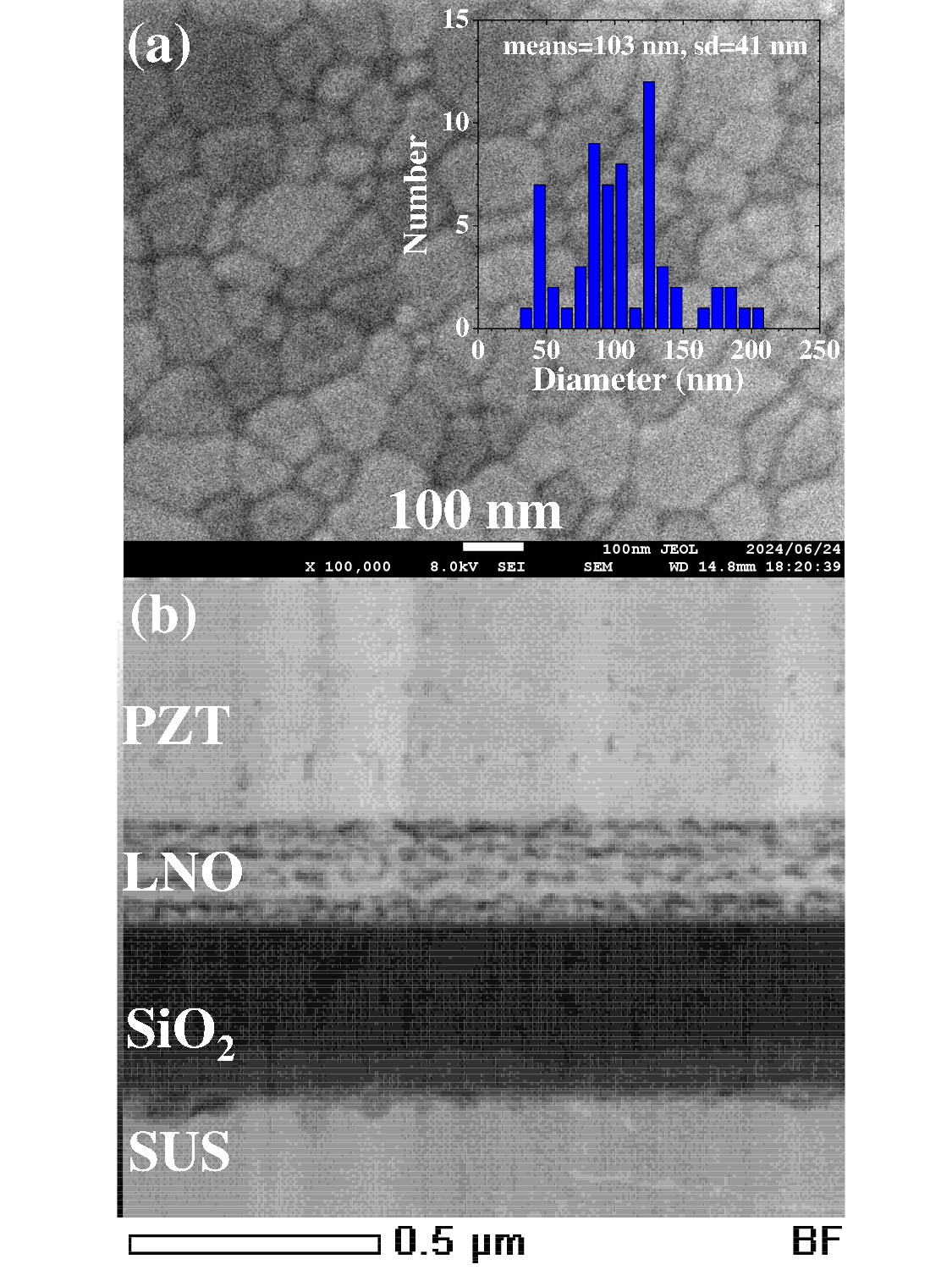}
\caption{\label{Fig3}  (a) Surface and  (b) cross-sectional images of   $c$-axis
oriented Pb(Zr$_{0.53}$Ti$_{0.47})$O$_3$ film. The distribution of the columnar PZT diameter is also shown in the inset of Figure  (a). }
\end{figure}

Figure \ref{Fig3} shows the surface and cross-sectional morphology of the $c$-axis oriented PZT film. From the cross-sectional view, it is evident that the growth mechanism of the PZT thin film is completely different from that of the LNO film. The LNO film appears to have experienced granular growth, whereas the PZT has grown in a columnar fashion. The average diameter of the columnar PZT is about  103 nm,  with a variation of 41 nm when observed from the surface. Since the columnar interfaces are not visible in the cross-section, it can be inferred that these columnar PZT crystals are interconnected. The surface of the PZT film is relatively smooth, with an average roughness of 0.88 nm as observed by AFM, and the height variation is within a few nanometers. An interesting phenomenon observed on the surface is that some areas in the SEM image appear darker while others are brighter, with both the bright and dark areas displaying uniform brightness. This FE-SEM observation was made without coating the surface with a metallic film. Since the brightness variation reflects the surface's charge state, it is reasonable to consider that the appearance of bright and dark areas is due to the emergence of 180-degree domains with opposite polarization directions. The observations from FE-SEM are consistent with the previously described X-ray results and the subsequent  polarization hysteresis and piezoelectric measurement results.

\begin{figure}
\includegraphics[width=8cm]{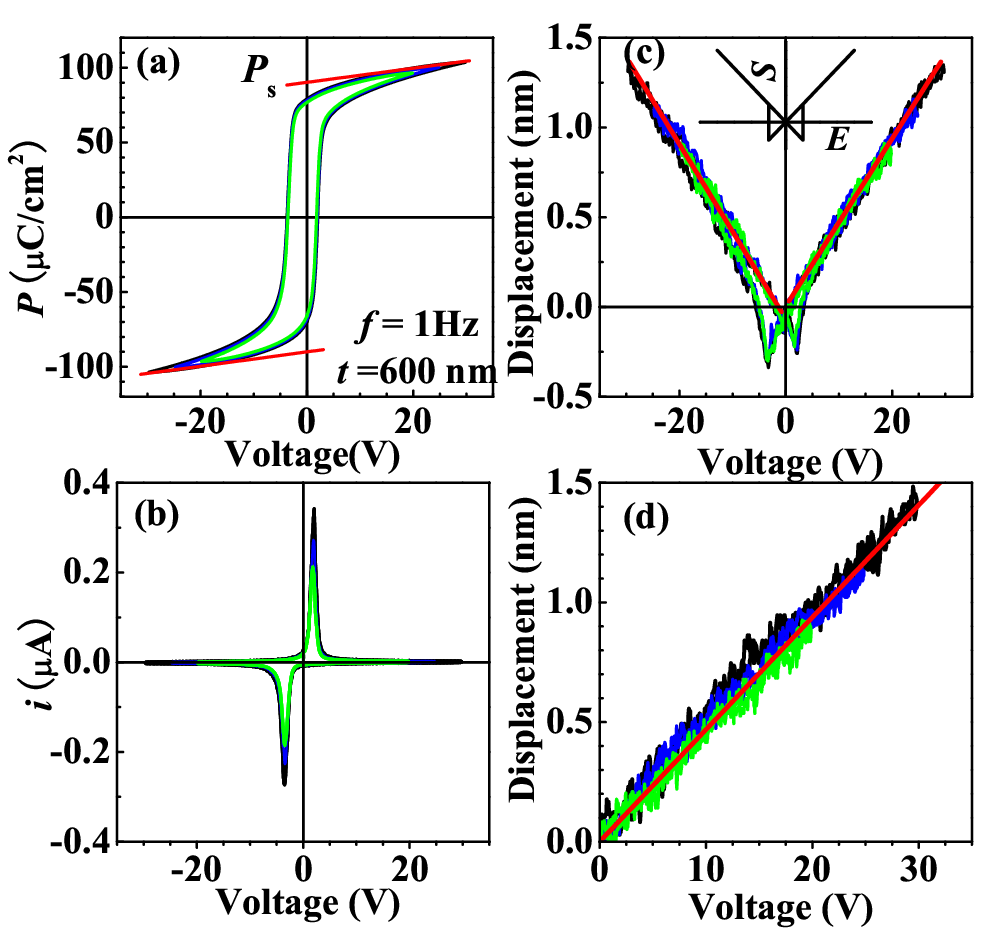}
\caption{\label{Fig4}  (a) $D-E$ hysteresis loop, (b) the switching
current and (c) the electric-field-induced  displacements in   $c$-axis
oriented Pb(Zr$_{0.53}$Ti$_{0.47})$O$_3$ under  bipolar voltage application.  (d)
Unipolar field-induced displacements. Inset in Fig.\ref{Fig3}(c) schematically shows an ideal  strain-field  response  in a crystal, where spontaneous  polarizations  reverse only by 180$^\circ$.\cite{Damjanovic} }
\end{figure}

Figure \ref{Fig4} illustrates typical results of the $D-E$ hysteresis loop, the switching current, and the electric-field-induced displacement for various applied bipolar or unipolar voltages using 600-nm thick PZT samples as an example. For the $c$-axis oriented PZT with tetragonal symmetry, only the 180-degree domain switching is anticipated when the sample is subjected to an electric field in the direction of spontaneous polarization. When the applied voltage exceeds the coercive field greatly, the switching current nearly drops to zero, and the electrical polarization reaches saturation. This indicates that the spontaneous polarization has been almost entirely switched to the field direction under such a high electric field in the $c$-axis oriented PZT sample. This is further corroborated by the displacement response of the sample obtained at the same time (Fig.\ref{Fig4}(c)). The $c$-axis oriented PZT sample exhibits the typical strain response behavior predicted for the single crystal, as shown in the inset of Fig.\ref{Fig4}(c).\cite{Damjanovic}  A linear piezoelectric response has been observed in the sample when removing the applied voltage, indicating almost complete  alignment of spontaneous polarization with  the field direction. For the case of unipolar field measurements, the displacement was measured after poling the sample. Once again, we have observed the linear piezoelectric effects as anticipated from the lattice response in the unipolar field measurements. The above results clearly indicate that the spontaneous polarization in $c$-axis oriented PZT has been aligned with the field direction. Under this state, one can use a linear fitting to  estimate the intrinsic piezoelectric coefficient $d_{33}$  of PZT, in which the extrinsic effects such as the domain effects are essentially negligible.

\begin{figure}
\includegraphics[width=8cm]{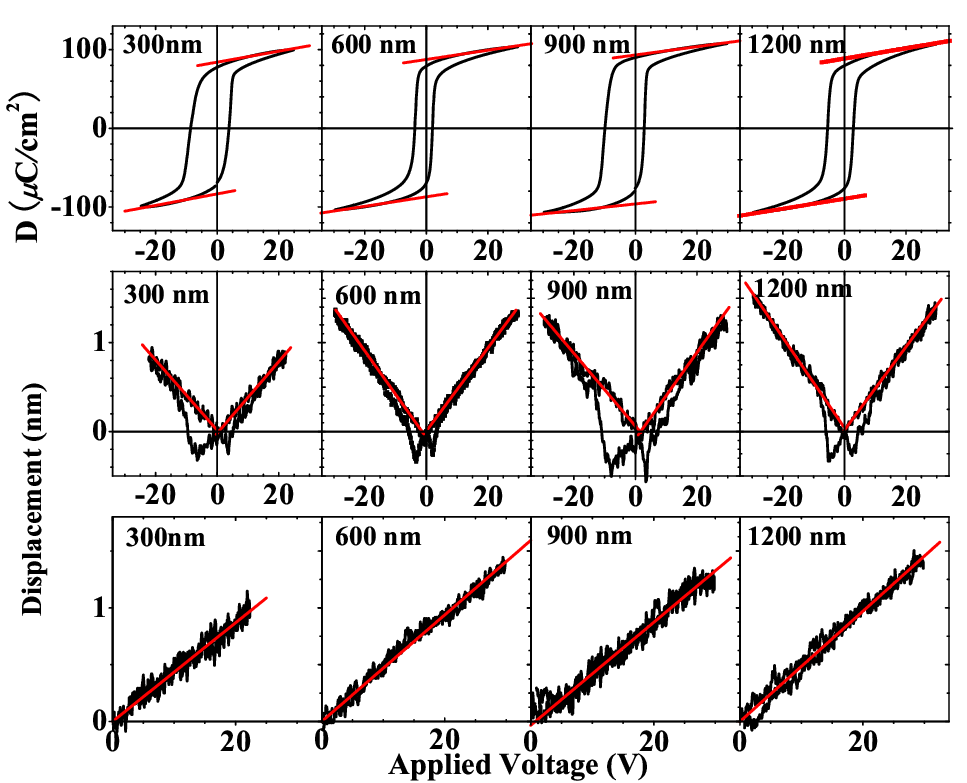}
\caption{\label{Fig5} $D-E$ hysteresis loop, bipolar  and unipolar field-induced displacements in $c$-axis oriented Pb(Zr$_{0.53}$Ti$_{0.47})$O$_3$  films with thickness ranging from 300 nm to 1200 nm. }
\end{figure}

\begin{figure}
\includegraphics[width=8cm]{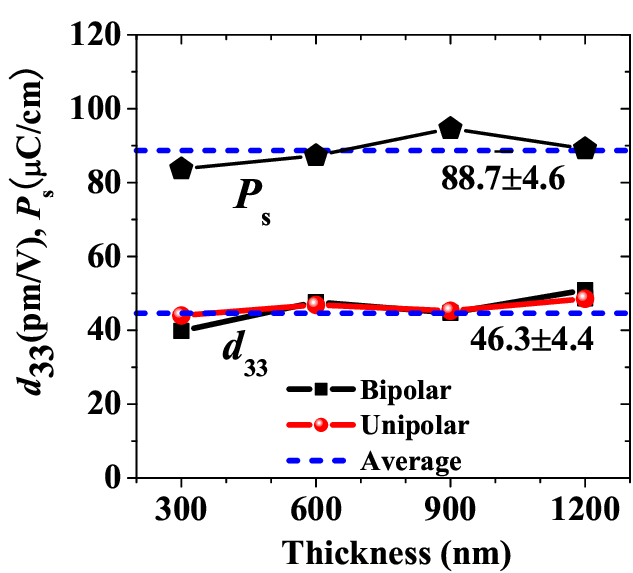}
\caption{\label{Fig6} The estimated $d_{33}$ values and spontaneous polarization $P_{\rm s}$ for  $c$-axis oriented Pb(Zr$_{0.53}$Ti$_{0.47})$O$_3$  films with thickness ranging from 300 nm to 1200 nm.}
\end{figure}

Fig.\ref{Fig5} summarizes the results of the $D-E$ hysteresis loop and the electric-field-induced displacement of the $c$-axis oriented PZT shown in Fig.\ref{Fig2}.  All samples demonstrate similar polarization switching and displacement behaviors. We then use the linear change in  displacement with the applied voltage to calculate the piezoelectric coefficient $d_{33}$  of these $c$-axis oriented PZTs by  linear fitting.  All samples show   similar  results for the piezoelectric coefficient,  as depicted in Fig.\ref{Fig6}. The $d_{33}$  was estimated to be  46.4 $\pm$ 4.4 pm/V from the statistical calculation of all the results obtained from  both the bipolar and unipolar displacement measurements of all samples.  Our experimental $d_{33}$ aligns well  with the  first-principles prediction   for PZT single crystal near MPB.\cite{Bellaiche2}  Using a first-principles-derived approach, Bellaiche et  al. studied the  finite-temperature behaviors of PZT near the  MPB  and demonstrated that the $d_{33}$ of single-crystal PZT with $x=0.5$  is around 50-55 pC/N at room temperature. However,  our experimental $d_{33}$ result disagrees  with the prediction  from the phenomenological Landau Devonshire theory by Du et al.,\cite{Du}   who  predicted   a $d_{33}$ value of 325 pm/V  for the $c$-axis  tetragonal PZT single crystal with $x=0.5$,   about 7 times larger  than our value.  This overestimated $d_{33}$ is very likely  caused by the accuracy of  the  Landau free-energy coefficients used in the calculation.\cite{Haun}  Unlike first-principles calculations, which offer insights at the atomic scale, Landau phenomenology provides a macroscopic understanding through experimental fitting parameters and cannot capture microscopic details. Moreover, the fitting parameters used for free-energy calculations are obtained from ceramics and are different from those of single crystals, leading to an incorrect estimation of the piezoelectric coefficients of the PZT single crystal.

The first-principles-derived calculation by Bellaiche et al. also  predicted  a  spontaneous polarization of 79 $\mu$C/cm$^2$ for this PZT single crystal near the  MPB.\cite{Bellaiche2} This prediction aligns well with our experimental statistical  result of 88.7 $\pm$ 4.6$\mu$ C/cm$^2$ as shown in Fig.\ref{Fig6}.   In our analysis, we  estimate the spontaneous polarization ($P_{\rm S}$)  from the  $D-E$ hysteresis loop  by assuming  that electrical displacement density $D$ is proportional to the  electric field $E$  when all spontaneous polarizations have been aligned along the field direction and make an extrapolation to  obtain  the spontaneous polarization  $P_{\rm s}$  at zero field,  as shown in  Fig.\ref{Fig4}(a) and Fig.\ref{Fig5}.

\begin{figure}
\includegraphics[width=8cm]{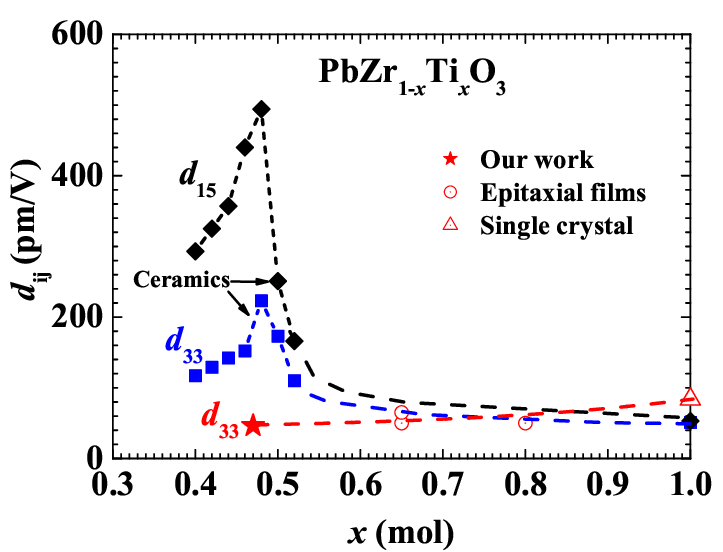}
\caption{\label{Fig7} Analysis of the  piezoelectric coefficients  of Pb(Zr$_{1-x}$Ti$_{x})$O$_3$\cite{Berlincourt} and PbTiO$_3$ (PT) \cite{Ikegami}  ceramics  in comparison to those of  reported epitaxial PZT  films\cite{Lee,Fujisawa,Grigoriev}, alongside the  polar-axis-oriented  Pb(Zr$_{0.53}$Ti$_{0.47})$O$_3$ films of this study and PT single crystal\cite{Li}. }
\end{figure}

Our $d_{33}$  of PZT with $x=0.47$ around MPB  compares  well with those  reported for   epitaxial PZT films  with higher PbTiO$_3$ content  and larger tetragonality $c/a$.\cite{Lee,Fujisawa,Grigoriev} The lattice distortion caused by an electric field has been shown to result in $d_{33}$ values of 50 pm/V for the  epitaxial 250-nm-thick  PZT film with $x=0.65$   and $c/a=4.133 \AA/3.989 \AA=1.036$,\cite{Lee}   65 pm/V for the epitaxial 2-$\mu$m-thick  PZT film with $x=0.65$   and $c/a=4.152 \AA/3.991 \AA=1.040$,\cite{Fujisawa}  and 45 pm/V for the epitaxial 35-nm-thick  PZT film with $x=0.8$   and $c/a=4.25 \AA/3.905  \AA=1.088$.\cite{Grigoriev}  Our result  and those reported values for epitaxial PZT are rather comparable to the  $d_{33}$ value  of  83.7 pm/V reported for PT single crystal.\cite{Li} The above comparison is also summarized in Fig.\ref{Fig7}. All these experimental findings confirm the   prediction derived from first-principles calculations that the piezoelectric  coefficients of tetragonal PZT around MPB are rather comparable to those of the simple PbTiO$_3$.\cite{Szabo,Bellaiche1,Bellaiche3,Bellaiche4}  For example, Bellaiche et al. predicted that the tetragonal PZT with $x=0.5$ and PbTiO$_3$   would have  $e_{33}$ piezoelectric coefficients of  3.4 C/m$^2$ and 3.8 C/m$^2$, respectively,  and thus concluded that "alloying PbTiO$_{3}$ with PbZrO$_3$ does not provide any enhancement of piezoelectricity with respect to PT".\cite{Bellaiche1}  This conclusion derived from  first-principles calculations is  supported by our experimental result for PZT near MPB and  those reported for PT single crystal and  epitaxial PZT single-crystalline films with different PT concentrations. 

In contrast, PZT ceramics exhibit different piezoelectric behaviors. In ceramics, the  piezoelectric response is dependent on the PT concentration,  and a maximum occurs around the MPB with $x=0.48$.\cite{Berlincourt}   PZT ceramics  near the MPB  with $x=0.48$  show a  $d_{33} $ piezoelectric coefficient  of  223 pm/V,\cite{Berlincourt}   which  is 4$\sim$5 times larger than  the 51 pm/V reported for PT ceramics\cite{Ikegami} and  our own measurements of the polar  $c$-axis $d_{33}$ of tetragonal PZT with  an MPB composition of  $x$=0.47.\cite{Berlincourt,Noheda2} For a comparison, the ceramics values reported for PZT  by Belincourt et al.\cite{Berlincourt}   and for PT by Ikegami et al.\cite{Ikegami}   are also replotted in Fig.\ref{Fig7}.

The significant $d_{33}$ piezoelectric coefficient observed in PZT ceramics near the  MPB cannot be solely attributed to the intrinsic  $d_{33}$ piezoelectric effect of the single crystal. Instead, it is largely influenced by extrinsic effects. A key extrinsic factor is the movement of domain walls or domain switching within the PZT ceramics.\cite{Damjanovic,Hall3,JLi,Gharb,Sun,Otonicar,Gorfman}

Near the MPB, PZT ceramics exhibit a complex domain structure due to the coexistence of tetragonal, rhombohedral, and monoclinic phases.\cite{Noheda2}  This complexity is further accentuated by the presence of nano-domains, averaging around 10 nm in size.\cite{Asada}  The high density and mobility of these nano-domain walls contribute substantially to the overall piezoelectric response ($d_{33}$), introducing nonlinear contributions beyond the intrinsic lattice strain.

Additionally, the coexistence of tetragonal and rhombohedral phases, interconnected by the monoclinic phase near the MPB,\cite{Noheda2}  facilitates interactions and local transformations between these phases under an applied electric field.\cite{Liu,Fan2,Zhao}  This phase transformation mechanism allows the ceramics to accommodate larger strains, thereby enhancing the piezoelectric response.

In polycrystalline PZT ceramics, grain boundaries also play a critical role in influencing piezoelectric behavior. These boundaries can act as sites for stress concentration and facilitate additional strain mechanisms that are not present in single-crystal materials.Unlike free-standing single crystal, the grains in ceramics are mechanically constrained by their neighboring grains. This  mechanical  constraint results in significant elastic strain during non-180° domain switching. Such intergranular and intragranular strains \cite{Hall1,Hall2,Sun}  can substantially enhance the piezoelectric response in PZT ceramics.

The significant $d_{15}$ effects predicted in PZT from first-principles calculations seem to provide a plausible and reliable explanation for the large  $d_{33}$ piezoelectric effect in PZT ceramics near the MPB. Using the predicted $d_{15}$ value of 580 pm/V, Bellaiche et al. successfully estimated a value of 163 pV/m for PZT ceramics with $x=0.5$ near the MPB at room temperature, \cite{Bellaiche2}   which closely matches the reported value of 173  pm/V for PZT ceramics with $x=0.5$.\cite{Berlincourt}   The large shear piezoelectric coefficient $d_{15}$ predicted for PZT crystal is consistent with the experimental results reported for ceramics, as shown in Fig. \ref{Fig7}, which indicate a $d_{15}$ of 494 pm/V for PZT ceramics near the MPB\cite{Berlincourt}   but only 53 pm/V for PT ceramics.\cite{Ikegami} However, further confirmation is needed for the single crystal.

\section{Conclusions}
In summary, we have presented  a  straightforward method to uncover  the intrinsic piezoelectric behaviors of PZT near the MPB along the polar axis. We found that  PZT with $x=0.47$  exhibits a $d_{33}$ piezoelectric coefficient of  46.4 $\pm$ 4.4 pm/V and a spontaneous  polarization of 88.7 $\pm$ 4.6
$\mu$C/cm$^2$  along the polar $c$-axis of the tetragonal structure. Our experimental results align well with predictions from first-principles-derived calculations but contradict with those  from phenomenological calculations. Our findings indicate that the significant piezoelectric effect ($d_{33}$) observed in PZT ceramics near the MPB is primarily driven by extrinsic effects, rather than the crystal’s intrinsic piezoelectric properties. These extrinsic effects include enhanced domain wall mobility, phase coexistence and transformation, and intergranular and intragranular strain effects. These mechanisms play a crucial role in amplifying the piezoelectric response, particularly in polycrystalline PZT ceramics. Understanding these extrinsic factors is essential for optimizing the piezoelectric performance of PZT-based materials for various applications. In technical applications of PZT films, achieving a substantial piezoelectric response requires careful consideration of these extrinsic effects, including domain reversal and movement.

Supporting Information: The process for preparing LaNiO$_3$(LNO) conducting layer, SiO$_2$ buffer layer,  and  PbZr$_{0.53}$Ti$_{0.47}$O$_3$ films (PDF).



\begin{acknowledgments}
 We are immensely grateful to Associate Professor Naonori Sakamoto from the Faculty of Engineering at Shizuoka University for his technical support in acquiring transmission electron microscopy images.
\end{acknowledgments}



\section{Reference}
\begin{thebibliography}{90}

\bibitem{Berlincourt} Berlincourt, D.;   Cmolik, C.;  Jaffe, H.  Piezoelectric properties of polycrystalline lead titanate zirconate compositions, Proc. IRE \textbf{1960}, {\it 48}, 220-229. DOI: 10.1109/JRPROC.1960.287467.

\bibitem{Jaffe1} Jaffe, B.;  Roth, R. S.;  Marzullo, S.  Piezoelectric properties of lead zirconate-lead titanate solid solution ceramics, J. Appl. Phys. \textbf{1954},  {\it 25}, 809-810. DOI:10.1063/1.1721741.

\bibitem{Jaffe2} Jaffe, B.;  Cook, W. R.;    Jaffe, H. {\it Piezoelectric Ceramics},  Academic: London, 1971.

\bibitem{Uchino} Uchino, K.  {\it Piezoelectric Actuators and Ultrasonic Motors},  Kluwer Academic Publishers: Boston, 1996.

\bibitem{Park} Park, S.-E.;  Shrout, T.E. Ultrahigh strain and piezoelectric behavior in relaxor based ferroelectric single crystals, J. Appl. Phys. \textbf{1997}, {\it 82}, 1804-1811. DOI:10.1063/1.365983.

\bibitem{Saito} Saito, Y.;  Takao, H.; Tani, T. ;  Nonoyama, T.;  Takatori, K.;   Homma, T.;  Nagaya, T.;    Nakamura, M. Lead-free piezoceramics, Nature \textbf{2004}, {\it 432}, 84-87. DOI:10.1038/nature03028.



\bibitem{Yang} Yang, S.; Bao, H.;  Zhou, C.; Wang, Y.; Ren, X;  Matsushita, Y;  Katsuya, Y.; Tanaka, M.;  Kobayashi, K.;  Song, X.; Gao, J. Large magnetostriction from morphotropic phase boundary in ferromagnets, Phys. Rev. Lett. \textbf{2010}, {\it 104}, 197201. DOI:10.1103/PhysRevLett.104.197201.

\bibitem{Bergstrom} 
Bergstrom, R.; Wuttig, M.;  Cullen, J.;  Zavalij, P.;  Briber, R.;  Dennis,  C.;
 Ovidiu Garlea, V.;   Laver M. Morphotropic phase boundaries in ferromagnets:Tb$_{1-x}$Dy$_x$Fe$_2$ Alloys, Phys. Rev. Lett. \textbf{2013}, {\it 111}, 017203. DOI:10.1103/PhysRevLett.111.017203.

\bibitem{George} George, A. M.;   Iniguez, J.;   Bellaiche, L.; Effects of atomic short-range order on the properties of perovskite alloys in their morphotropic phase boundary, Phys. Rev. Lett. \textbf{2003}, {\it 91}, 045504. DOI:10.1103/PhysRevLett.91.045504.

\bibitem{Fu}  Fu, H.; Cohen, R. E.  Polarization rotation mechanism for ultrahigh electromechanical response in single-crystal piezoelectrics, Nature
\textbf{2000}, {\it 403}, 281-283. DOI:10.1038/35002022.

\bibitem{Guo} Guo, R.;  Cross, L. E.;   Park, S-E.;  Noheda, B.; Cox, D. E.;  Shirane, G. Origin of the high piezoelectric response in PbZr$_{1-x}$Ti$_x$O$_3$, Phys. Rev. Lett. \textbf{2000}, {\it 84}, 5423-5426. DOI:10.1103/PhysRevLett.84.5423.

\bibitem{Xu} Xu, G.;  Wen, J.;  Stock, C.;   Gehring, P. M.;  Phase instability induced by polar nanoregions in a relaxor ferroelectric system, Nature materials \textbf{2008}, {\it 7}, 562-566. DOI:10.1038/nmat2196.

\bibitem{Noheda1}  Noheda, B.; Cox, D.; Shirane, G.; Gonzalo, J.; Cross, L.;   Park, S.  A monoclinic ferroelectric phase in the Pb(Zr$_{1-x}$Ti$_x$)O$_3$ solid solution, Appl. Phys. Lett. \textbf{1999}, {\it 74}, 13089. DOI: 10.1063/1.123756.

\bibitem{Noheda2} Noheda B.;  Cox, D. E., Shirane, G.;  Guo R.;  Jones B.;  L. E. Cross, L.E., Stability of the monoclinic phase in the ferroelectric perovskite PbZr$_{1-x}$Ti$_x$O$_3$, Phys. Rev. B \textbf{2000}, {\it 63},  014103. DOI: 10.1103/PhysRevB.63.014103.

\bibitem{Hall1} Hall, D. A.; Steuwer, A.; Cherdhirunkorn, B.; Mori, T.; Withers  P. J.  A high energy synchrotron x-ray study of crystallographic texture and lattice strain in soft lead zirconate titanate ceramics
J. App. Phys.  \textbf{2004}, {\it 96}, 4245-4252. DOI: 10.1063/1.1787590.

\bibitem{Hall2} Hall, D. A.; Steuwer, A.; Cherdhirunkorn, B.; Mori, T.; Withers  P. J.  Analysis of elastic strain and crystallographic texture in poled rhombohedral PZT ceramics, Acta Materialia  \textbf{2006}, {\it 54}, 3075-3083. DOI: 10.1016/j.actamat.2006.02.043

\bibitem{Gharb} Bassiri-Gharb, N;   Fujii, I;  Hong, E.; Trolier-McKinstry S.;  Taylor, D. V.; Damjanovic
D. Domain wall contributions to the properties of piezoelectric thin films, J. Electroceram. \textbf{2007}, {\it 19}, 47-65: DOI:10.1007/s10832-007-9001.

\bibitem{Hinterstein} Hinterstein, M.; Rouquette, J.;  Haines, J.; Papet, Ph.; Knapp, M.; Glaum, J.;  Fuess, H. 
Structural description of the macroscopic piezo- and ferroelectric properties
 of lead zirconate titanate,  Phys. Rev. Lett. \textbf{2011}, {\it 107}, 077602. DOI: 10.1103/PhysRevLett.107.077602.
 
\bibitem{Liu} Liu, H.; Chen, J.;  Huang, H.;  Fan, L.; Ren, Y.; Pan, Z.;   Deng, J.;   Chen, L-Q.; Xing, X.  Role of reversible phase transformation for strong piezoelectric performance at the morphotropic phase boundary, Phys. Rev. Lett. \textbf{2018}, {\it 120}, 055501.DOI: 10.1103/PhysRevLett.120.055501
 
  
\bibitem{Fan1} Fan, L.;  Chen, J.;  Ren Y.;  Pan, Z.; Zhang, L.; Xing X. Unique piezoelectric properties of the monoclinic phase in Pb(Zr,Ti)O$_3$  ceramics: large lattice strain and negligible domain switching,  Phys. Rev. Lett. \textbf{2016}, {\it 116}, 02760. DOI:10.1103/PhysRevLett.116.027601.

 \bibitem{Fan2} Fan, L.;  Zhang, L.; Liu,  H. Direct observation of polarization rotation in the monoclinic
  $M_{\rm B}$ phase under electrical loading,  Inorg. Chem. \textbf{2021}, {\it 60}, 15190-15195 . DOI:10.1021/acs.inorgchem.1c01599. 
 
 \bibitem{Wang}  Wang, Z.; Zhang, N.; Yokota, H.; Glazer, A. M.; Yoneda, Y.; Ren, W.; Ye, Z-G.
 Local structures and temperature-driven polarization rotation in Zr-rich PbZr$_{1-x}$Ti$_x$O$_3$, Appl. Phys. Lett. \textbf{2018}, {\it 113}, 012901. DOI:10.1063/1.5024422.

 \bibitem{Zhao} Zhao, J.; Funni, S. D.; Molina, E. R.; Dickey, D. C.; Jones, J. L. Orientation-dependent, field-induced phase transitions in soft lead zirconate titanate piezoceramics, J. Euro. Ceram. Soc.  \textbf{2021}, {\it 41}, 3357-3362.  DOI:10.1016/j.jeurceramsoc.2021.01.043.
 
\bibitem{Sun} Sun, S.; Zhang, Y.; Fan, L.; Deng, S.; Gao, B.; Ren, Y.; Liu, H.; Chen, J. Role of tetragonal distortion on domain switching and lattice strain of piezoelectrics by in-situ synchrotron diffraction, Scripta Materialia \textbf{2021}, {\it194}, 113627. DOI:10.1016/j.scriptamat.2020.11.012.

\bibitem{JLi} Li, J.; Rogan, R.; Ustundag, E.; Bhattacharya, K. Domain switching in polycrystalline ferroelectric ceramics. Nature Mater \textbf{2005}, {\it4}, 776-781. DOI:10.1038/nmat1485


\bibitem{Asada} Asada, T;   Koyama, Y.  Ferroelectric domain structures around the morphotropic phase boundary of the piezoelectric material, Phys. Rev. B, \textbf{2007}, {\it 75}, 214111. DOI: 10.1103/PhysRevB.75.214111.


\bibitem{Otonicar} Otonicar, M.; Dragomir M.; Rojac, T. Dynamics of domain walls in ferroelectrics and relaxors, J. Am. Ceram. Soc. \textbf{2022}, {\it 105}, 6479-6507. DOI: 10.1111/jace.18623.



\bibitem{Gorfman} Gorfman, S.; Bokov, A. A.; Davtyan, A.; Reiser, M.;  Xie, Y.;  Ye, Z-G.; Zozulya,  A. V.; Sprung,M.;  Pietsch, U.;  Gutt, C. Ferroelectric domain wall dynamics characterized with X-ray photon correlation spectroscopy, PNAS, \textbf{2018}, {\it 115},  E6680-E6689. DOI:10.1073/pnas.1720991115.

\bibitem{Shirane} Shirane, G.;  Suzuki, K.   Crystal structure of Pb(Zr-Ti)O$_3$, J. Phys. Soc. Jpn. \textbf{1952}, {\it 7}, 333.  DOI:10.1143/JPSJ.7.333.

\bibitem{Clarke} Clarke, R.;  Whatmore,  R. W.;  Glazer, A. M.  Growth and characterization of PbZr$_x$Ti$_{1-x}$O$_3$ single crystals, Ferroelectrics,  \textbf{1976}, {\it 13}, 497-500. DOI:10.1080/00150197608236650.
%

\bibitem{Phelan} Phelan, D.;  Long, X.; Xie, Y.;  Ye, Z.-G.;  Glazer, A. M.; Yokota, H.; 
Thomas, P. A.; Gehring, P. M. Single crystal study of competing rhombohedral and monoclinic order in lead zirconate titanate, Phys. Rev. Lett. \textbf{2010}, {\it 105}, 207601. DOI: 10.1103/PhysRevLett.105.207601.

\bibitem{Bokov} Bokov, A. A.; Long, X.;   Ye, Z.-G.  Optically isotropic and monoclinic ferroelectric phases in Pb(Zr$_{1-x}$Ti$_x$)O$_3$(PZT) single crystals near morphotropic phase boundary, Phys. Rev. B  \textbf{2010}, {\it 81}, 172103. DOI:10.1103/PhysRevB.81.172103. 

\bibitem{Roleder} Roleder, K.;   Majchrowski, A.; Lazar, I.;  Whatmore, R. W.;  Glazer, A. M.;   Kajewski, D.;   Koperski, J.;  Soszyski, A. Monoclinic domain populations and enhancement of piezoelectric properties in a PZT single crystal at the morphotropic phase boundary, Phys. Rev. B  \textbf{2022}, {\it 105}, 144104. DOI:10.1103/PhysRevB.105.144104.

\bibitem{Whatmore} Lazar, I.; Whatmore, R. W.; Majchrowski, A.;  Glazer, A. M.;  Kajewski, D.; Koperski, J.; Soszynski,  A.; Piecha, J.;  Loska, B.;  Roleder, K. Ultrahigh Piezoelectric Strains in PbZr$_{1-x}$Ti$_x$O$_3$ Single crystals with controlled Ti content close to the tricritical point, Materials \textbf{2022}, {\it 15}, 6708. DOI:10.3390/ma15196708.


\bibitem{Du}Du, X.-H.; Zheng, J.;   Belegundu, U.;  Uchino, K. Crystal orientation dependence of piezoelectric properties of lead zirconate titanate near the morphotropic phase boundary, Appl. Phys. Lett. \textbf{1987}, {\it 72},  2421-2423. DOI:10.1063/1.121373.

\bibitem{Szabo} Saghi-Szabo, G.;   Cohen, R. E.; Krakauer,  H.  First-principles study of piezoelectricity in tetragonal PbTiO$_3$ and PbZr$_{1/2}$Ti$_{1/2}$O$_3$, Phys. Rev. B \textbf{1999}, {\it 83}, 12771-12776. DOI:10.1103/PhysRevB.59.12771.

\bibitem{Bellaiche1} Bellaiche, L.; Vanderbilt, D. Intrinsic piezoelectric response in perovskite alloys: PMN-PT versus PZT, Phys. Rev. Lett. \textbf{1999}, {\it 83}, 1347-1350. DOI:10.1103/PhysRevLett.83.1347.

\bibitem{Bellaiche2} Bellaiche, L.; Garca, Al.; Vanderbilt, D. Finite-temperature properties of Pb(Zr$_{1-x}$Ti$_x$)O$_3$ alloys from first principles, 
Phys. Rev. Lett. \textbf{2000}, {\it 84}, 5427-5430. DOI:10.1103/PhysRevLett.84.5427.


\bibitem{Suzuki} Suzuki, H.; Miwa, Y.; Naoe, T.;  Miyazaki, H.; Ota , T.;  Fuji, M.;   Takahashi, M.   Orientation control and electrical properties of PZT/LNO capacitor through chemical solution deposition,  J. Euro. Ceram. Soc.  \textbf{2006}, {\it 26},  1953-1956. DOI:10.1016/j.jeurceramsoc.2005.09.037.

\bibitem{Miyake} Miyake, S.;Yamamoto, K.;  Fujihara, S.; Kimura, T.  (100)-orientation of pseudocubic perovskite-type LaNiO$_3$ thin films on glass substrates via the Sol-Gel process, J. Am. Ceram. Soc.  \textbf{2002}, {\it 85}, 992-994. DOI: 10.1111/j.1151-2916.2002.tb00206.x.

\bibitem{Schlom} Schlom, D. G.; Chen, L.-Q.; 
 Eom, C.-B.;  Rabe, K. M.;  Streiffer, S. K.;  Triscone, J.-M.   Strain tuning of ferroelectric thin films, Annu. Rev. Mater. Res. \textbf{2007}, {\it 37}, 589-626. DOI: 10.1146/annurev.matsci.37.061206.113016.

\bibitem{Haun} Haun, M. J.;   Furman, E.;  Jang, S. J.;  Cross, L. E.  Thermodynamic theory of the lead zirconate-titanate solid solution system, part I: Phenomenology, Ferroelectrics \textbf{1989}, {\it 99}, 13-25. DOI:10.1080/00150198908221436.


\bibitem{Lee}  Lee ,H. J.;  Shimizu,T.;   Funakubo, H.;  Imai, Y.; Sakata, O.;  Hwang, S. H.; 
 Kim, T. Y.;  Yoon, C.;  Dai, C.; Chen, L. Q.;   Lee, S. Y.; Jo, J. Y. Electric-field-driven nanosecond ferroelastic-domain switching dynamics in epitaxial Pb(Zr,Ti)O$_3$ film, Phys. Rev. Lett. \textbf{2019}, {\it 123}, 217601. DOI:10.1103/PhysRevLett.123.217601.
 
\bibitem{Grigoriev} Grigoriev,  A.;   Sichel, R.; Lee, H. N.;  Landahl, E. C.; Adams, B.;  Dufresne, E. M.; Evans, P.G. Nonlinear piezoelectricity in epitaxial ferroelectrics at high electric fields, Phys. Rev. Lett. \textbf{2008}, {\it 100}, 027604. DOI:10.1103/PhysRevLett.100.027604 .


\bibitem{Fujisawa} Fujisawa, T.;  Ehara, Y.; Yasui, S.;  Kamo, T.; Yamada, T.;   Sakata, O.;  Funakubo, H. Direct observation of intrinsic piezoelectricity of Pb(Zr,Ti)O$_3$ by time-resolved x-ray diffraction measurement using single-crystalline films, Appl. Phys. Lett. \textbf{2014}, {\it 105}, 012905. DOI:10.1063/1.4889803.


\bibitem{Li}  Li, Z.; Grimsditch, M.;  Xu, X.;  Chan, S.-K. The elastic, piezoelectric and dielectric constants of tetragonal PbTiO$_3$ single crystals, Ferroelectrics \textbf{1993}, {\it 141}, 313-325. DOI: 10.1080/00150199308223459.

\bibitem{Bellaiche3} Bellaiche, L.;  Vanderbilt, D.  Virtual crystal approximation revisited: application to dielectric and piezoelectric properties of perovskites, Phys. Rev. B \textbf{2000}, {\it 61},  7877-7882. DOI: 10.1103/PhysRevB.61.7877.


\bibitem{Bellaiche4}  Bellaiche, L.  Piezoelectricity of ferroelectric perovskites from first principles, Current Opinion in Solid State and Materials Science  \textbf{2002}, {\it 6}, 19-25. DOI:10.1016/S1359-0286(02)00017-7.



\bibitem{Ikegami}  Ikegami, S.; Ueda, I.;  Nagata, T.  Electromechanical properties of PbTiO$_3$ ceramics containing La and Mn, J. Acoust. Soc. Am. \textbf{1971}, {\it 50}, 1060-1066. DOI:10.1121/1.1912729.

\bibitem{Damjanovic} Damjanovic, D. Ferroelectric, dielectric and piezoelectric properties of ferroelectric thin films and ceramics, Rep. Prog. Phys. \textbf{1998}, {\it 61}, 1267-1324. DOI:10.1088/0034-4885/61/9/002.

\bibitem{Hall3} Hall, D.A. Review nonlinearity in piezoelectric ceramics. J. Mater. Sci.  \textbf{2001}, {\it 36}, 4575-4601 (2001). https://doi.org/10.1023/A:1017959111402




 

 \end{thebibliography}

\clearpage




\clearpage \clearpage
\newpage

\newpage

\newpage

\newpage

\newpage

\newpage

\newpage

\end{document}